\documentclass[a4paper,10pt]{article}
\usepackage{optro2010}
\usepackage[latin1]{inputenc}
\usepackage[french,english]{babel}
\usepackage{graphicx,afterpage}
\usepackage{psboxit,pstricks,rotating}
\usepackage{ulem}
\usepackage{amsmath,bbm} 
\usepackage{amsfonts} 
\usepackage{amssymb}
\usepackage{vector}
\usepackage{theorem}
\usepackage{stmaryrd}
\usepackage{euscript}
\usepackage{pifont}
\usepackage{array}
\usepackage[T1]{fontenc}
\usepackage{endnotes}

\title{\large \textsf{METRIC: a complete methodology for performances evaluation of automatic target Detection, Recognition and Tracking algorithms in infrared imagery}}

\name{\normalsize\textsf{$^{1}$J\'er\^ome Gilles, $^{2}$St\'ephane Landeau, $^{2}$Tristan Dagobert,$^{3}$Philippe Chevalier, $^{4}$Eric Sti\'ee, $^{5}$Damien Diaz, $^{5}$Jean-Luc Maillart}}
\address{\normalsize\textit{$^1$DGA/CEP/EORD, 16bis rue Prieur de la C\^ote d'Or 94110 Arcueil France, jerome.gilles@dga.defense.gouv.fr,}\\
\normalsize\textit{$^2$DGA/CEP/LOT, 16bis rue Prieur de la C\^ote d'Or 94110 Arcueil France, stephane.landeau@dga.defense.gouv.fr, tristan.dagobert@dga.defense.gouv.fr,}\\
\normalsize\textit{$^3$DGA/GA-ETAS, BP 60036, Montreuil-Juign\'e 49245 Avrill\'e Cedex France, philippe.chevalier@dga.defense.gouv.fr,}\\
\normalsize\textit{$^4$DGA/CEP/CGN/OP, 7-9 rue des Mathurins 92221 Bagneux Cedex France, eric.stiee@dga.defense.gouv.fr,}\\
\normalsize\textit{$^5$Bertin Technologies, D\'epartement Bertin Syst\`emes - P\^ole Architecture Syst\`emes, 155 rue Louis Armand, P\^ole d'activit\'e d'Aix en Provence, BP22000, 13791 Aix en Provence France, diaz@bertin.fr, maillart@bertin.fr}
}

\begin{document}

\maketitle

\begin{abstract}
In this communication, we deal with the question of automatic target detection, recognition and tracking (ATD/R/T) algorithms performance assessment. We propose a complete methodology of evaluation which approaches objective image datasets development and adapted metrics definition for the different tasks (detection, recognition and tracking). We present some performance results which are currently processed in a French-MoD program called 2ACI (``Acquisition Automatique de Cibles par Imagerie``).
\end{abstract}
\sffamily

\section{Introduction}
Today, image processing is more and more used in military applications. Ones of the most used are Automatic Target Detection/Recognition  (ATD/R) and tra\-cking algorithms on infrared imagery. An objective of the french governmental agency DGA is assessment of such kind of algorithms in order to evaluate their performances.\\

\begin{figure}[!h]
\centering\includegraphics[width=\columnwidth]{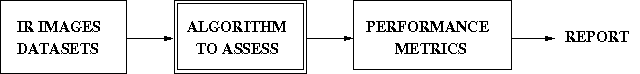}
\caption{Principle of ATD/R algorithm evaluation.}
\label{fig:evalconcept}
\end{figure}

An important part of work in a french program called 2ACI (``Acquisition Automatique de Cibles par Imagerie'') is the definition of a complete performance evaluation methodology. The performance evaluation principle is given in figure \ref{fig:evalconcept}. We start with the definition of how to build test image datasets. Another important part is: which metrics are relevant to measure performances of ATD/R and tracking algorithms? 

The paper is organized as follows. In section \ref{sec:metrics}, we present the different metrics we choose to assess the performances of ATD/R and tracking algorithms. Section \ref{sec:data} is devoted to the construction of image datasets. In section \ref{sec:test}, we present some preliminary results of the french 2ACI program. We end the paper by giving a conclusion and some perspectives of this work.

\section{Assessment metrics}\label{sec:metrics}
In this section, we adress the choice of relevant metrics to quantify the performances of the different kind of algorithms. As the different tasks has 
their own characteristics, we propose adapted metrics for each one. In all this section, we assume that test datasets and their ground truth are 
available.
\subsection{Detection}\label{sec:detect}
\begin{figure}[!h]
\centering\includegraphics[width=\columnwidth]{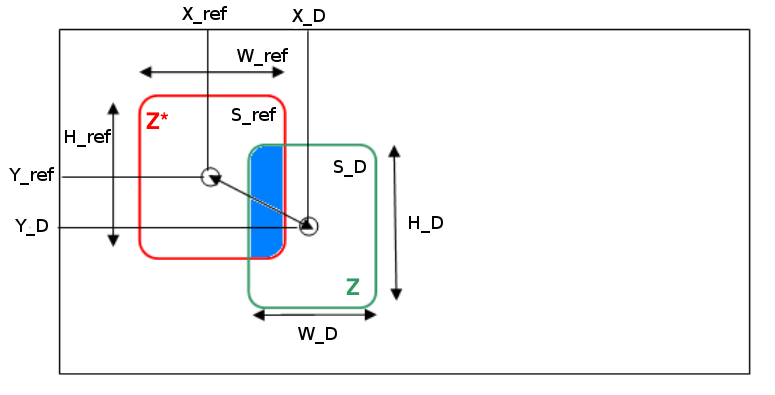}
\caption{Detection's notations.}
\label{fig:notation}
\end{figure}

The detection algorithm is said efficient when the target is detected, well localized and its size is well estimated. Let us define some notations based 
on figure \ref{fig:notation}. We assume that the assessed algorithm outputs the bounding box (BBox) around the detected target. The reference target is 
denoted $Z^*$ and the detected one by $Z$. The variables $X_{ref}, Y_{ref}, W_{ref}, H_{ref}$ and $S_{ref}$ are respectively the coordinates of the 
BBox's center, $W_{ref}$, $H_{ref}$ and $S_{ref}$ are the width, height and surface of the BBox of the reference target. The same definitions are used 
for the detected target ($D$ is the corresponding subscript).

In practice, we have two ways to declare a detection as good or not. The first one is to use Jaccard's criterion \cite{jaccard,jaccard2} (Eq.\ref{eq:jaccard}).
\begin{equation}\label{eq:jaccard}
m_0(Z,Z^*)=\frac{|\bigcap(Z,Z^*)|}{|\bigcup(Z,Z^*)|}>\epsilon_0=0.5
\end{equation}

The second one, inpired from the ROBIN competition \cite{robin}, is to use a combinaison of three criteria: a localization criterion $m_1$, a scale accuracy criterion $m_2$ and a segmentation accuracy criterion $m_3$, respectively defined by equations (\ref{eq:m1}),(\ref{eq:m2}),(\ref{eq:m3}).
\begin{align}
&m_1(Z,Z^*)= \label{eq:m1} \\ \notag
&\frac{2}{\pi}\arctan\left(\max\left(\frac{|X_D-X_{ref}|}{W_{ref}},\frac{|Y_D-Y_{ref}|}{H_{ref}}\right)\right) \\
&m_2(Z,Z^*)=\frac{|S_D-S_{ref}|}{\max(S_d,S_{ref})} \label{eq:m2} \\
&m_3(Z,Z^*)=\frac{2}{\pi}\arctan\left|\frac{H_D}{W_D}-\frac{H_{ref}}{W_{ref}}\right| \label{eq:m3}
\end{align}
Then a detection is said good if $m_1\leq\epsilon_1$, $m_2\leq\epsilon_2$ and $m_3\leq\epsilon_3$ are simultaneously verified, where we experimentaly choose $\epsilon_1=\epsilon_3=0.15$ and $\epsilon_2=0.5$. Finally, we can calculate, over the dataset, the (good) detection rate ($DR$), the false alarm rate ($FAR$) and then plot the corresponding ROC curve \cite{roccurve}.

\begin{figure}[!h]
\centering\includegraphics[width=\columnwidth]{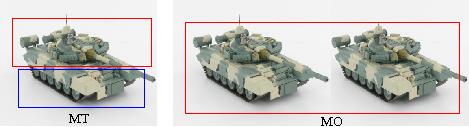}
\caption{Multiple tracker (MT) and multiple objects (MO) definition.}
\label{fig:smith1}
\end{figure}

In \cite{smith}, the authors propose two other interesting metrics which take care about another aspect of segmentation accuracy: the multiple trackers (MT) and multiple objects (MO). The first one represents the fact that multiple BBoxes are found on a unique target, the second one, the case of a unique BBox on multiple targets (see figure \ref{fig:smith1}).

All these metrics permit to accurately evaluate the behaviour and performances of any ATD algorithms.

\subsection{Recognition}
We select two levels of classification: recognition and identification. The first one uses general classes (car, truck, tank, $\ldots$). The classes used by the second one correspond to detailed model of target (AMX30, Leclerc, T72, $\ldots$). In order to evaluate the performances of this kind of algorithms, we need to check if the class proposed by the ATR algorithm is or not the same as the reference class. The best way to summarize these results is to use confusion matrices \cite{confmatrix}. 

\subsection{Tracking}
\begin{figure}[!h]
\centering\includegraphics[width=\columnwidth,height=4cm]{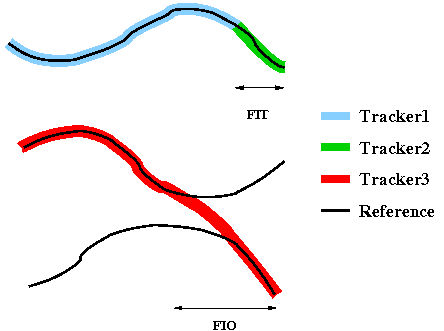}
\caption{False Identified Trackers and Objects.}
\label{fig:fitfio}
\end{figure}

In this section, we examine the case of tracking performed by movement detection algorithms. Two points are needed to be evaluated: the detection of target and the tracking itself. The detection case can be treated with the same metrics described in section \ref{sec:detect}. In this section, we specifically add some metrics to deal with the performances of tracking. In \cite{smith}, the authors adress the behavior of a tracker, in the sense that the algorithm could assign successive trackers to a same target or a tracker initially assigned to one target could ``jump'' to another target. The first one is called the False Identified Tracker (FIT) and the second one the False Identified Object (FIO). For example, in figure \ref{fig:fitfio}, the first target is associated with Tracker1 but at a certain time the algorithm missed this tracker and create a new tracker: Tracker2, this is FIT. Tracker3 is assigned to the second target but jumps to the third target at a certain time, this corresponds to FIO.\\

All the metrics described in the previous sections permit to accurately evaluate the behavior and performances of any ATD/R and tracking algorithms.

\section{Infrared image dataset generation}\label{sec:data}
The acquisition of IR databases presents a relatively high cost and takes a lot of time. A solution holds in the use of scenes simulators, but those remain expensive in computing time and it is especially difficult to select the various parameters in order to sweep a maximum of operational scenarios in an exhaustive way. In previous papers \cite{spie06,itbms09}, we presented an hybrid simulation method which permit to generate huge databases easily. We briefly recall this method in the next subsections.

\subsection{Hybrid IR image generation}
\begin{figure}[t]
\begin{center}
\includegraphics[scale=0.35]{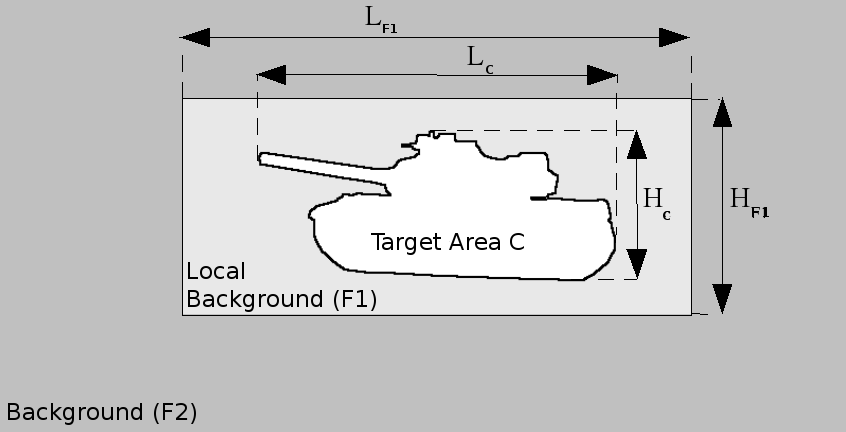}
\end{center}
\caption{Definition of different areas for target superimposition over a chosen background.}
\label{fig:zones}
\end{figure}

\begin{figure}[t]
\begin{center}
\includegraphics[width=0.46\textwidth]{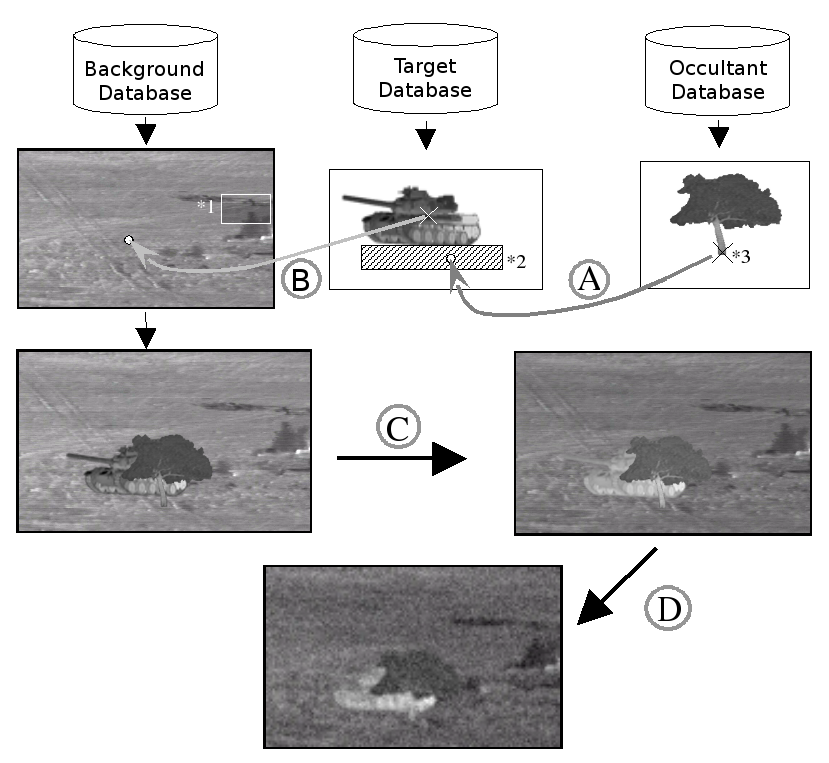}
\end{center}
\caption{Hybrid scene generation principle.}
\label{fig:gene}
\end{figure}

In \cite{spie06}, the authors propose to generate hybrid data\-bases by superimposition of targets and occultant (like trees, rocks,$\ldots$) in front of a background under constraint of image quality metrics. Entry parameters of these constraints are most effective to describe realistic operational scenarios. The used constraints are: local contrast $RSS$, ``detectability'' quantity $Q_D$, signal to clutter ratio $SCR$, occultation ratio $R_x$ and internal target contrast $K$. These quantities are defined by

\begin{align}
RSS&=\frac{1}{\nu_k}\sqrt{(\mu_{C}-\mu_{F_1})^2+\sigma_{C}^2} \\
Q_D&=RSS.S_{C} \\
SCR&=\frac{\nu_kRSS}{\sigma_F} \\
R_x&=\frac{S_{\text{occluded target area}}}{S_{\text{full target area}}} \\
K&=\frac{\mu_{F_1}-\mu_{C}}{\nu_kRSS}=\frac{\Delta\mu}{\nu_kRSS}
\end{align}

where $C$ is the target, $F_1$ the local background over $C$ and $F_2$ the remaining background (we denote the global background $F=F_1\cup F_2$), see Figure \ref{fig:zones}. The quantities $S_x$, $\mu_x$, $\sigma_x$ are the surface, mean and standard deviation of $x$ area where $x$ is $C$, $F_1$ or $F_2$, respectively. The coefficient $\nu_k$ is the coefficient which permits to do the conversion between pixel gray levels and temperature in Kelvin. The choice of these parameters fixes some gains and offsets to apply on the pixels of both the target and background in order to obtain the resulting image. Finally, the sensor effect (MTF, sampling and noise) is applied. The hybrid scene generation process is summarised in Figure \ref{fig:gene}. We start, \textcircled{\footnotesize{A}}, by positioning the occultant, then, \textcircled{\footnotesize{B}}, the positioning of the target inside the background. We apply the calculated gains and offsets to histograms of each region, \textcircled{\footnotesize{C}}. We finish by applying the sensor effect, \textcircled{\footnotesize{D}}. More details and the expressions of the different gains and offsets to apply can be found in \cite{spie06}. This scene generation principle was used for ATR algorithms evaluation in the CALADIOM project. 

However, one aspect is not taken into account in this algorithm: the intrinsic thermal variability of targets. In the next subsection, we propose an approach to deal with this aspect in the hybrid scene generation.

\subsection{Intrinsic thermal variability}
Here we recall an original method proposed in \cite{itbms09} allowing to take into account of this variability during the scene generation. 

\begin{figure}[t]
\begin{center}
\includegraphics[width=0.35\textwidth]{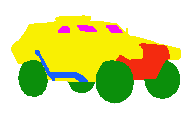}
\end{center}
\caption{Map of homogeneous thermal behavior of a given vehicle.}
\label{fig:decoup}
\end{figure}

This variability, which is function of the vehicle operation, is equivalent to a modification of the vehicle's signature. It is too complex, from a practical point of view, to use an accurate thermal physical models for different targets. We propose to create intermediate signatures by interpolation from ambient ($TA$) and operationnal ($TF$) temperatures, taken from real radiometric images from ETAS (a french center which does signatures measures). For this purpose, we lay out 3D models of vehicles on which we plate infrared textures. These textures are available for the $TA$ and $TF$ temperatures. We propose to segment the surface of the vehicle into homogeneous thermal behavior areas which are dependent on the different operational vehicle's areas. The relevant chosen areas are: the engine, the main body, the muffler, windows, tires/caterpillar (see Figure \ref{fig:decoup}).

An intermediate thermal state of an area $R$ ($TI_R$), relevant to the wanted variability, is generated by mixing the states $TA$ and $TF$, according to equation (\ref{eq:mix}).

\begin{equation}\label{eq:mix}
TI_R=(1-\lambda_R)TA_R+\lambda_R TF_R,
\end{equation}
where $\lambda_R\in [0;1]$ represent the variability rate for area $R$. We can define three different behavior:
\begin{enumerate}
\item ambient temperature: $\lambda_R\in[0;0.1]$,
\item intermediate temperature: $\lambda_R\in]0.1;0.9[$,
\item in operation temperature: $\lambda_R\in[0.9;1]$.
\end{enumerate}

\begin{figure}[t]
\begin{center}
\includegraphics[scale=0.2]{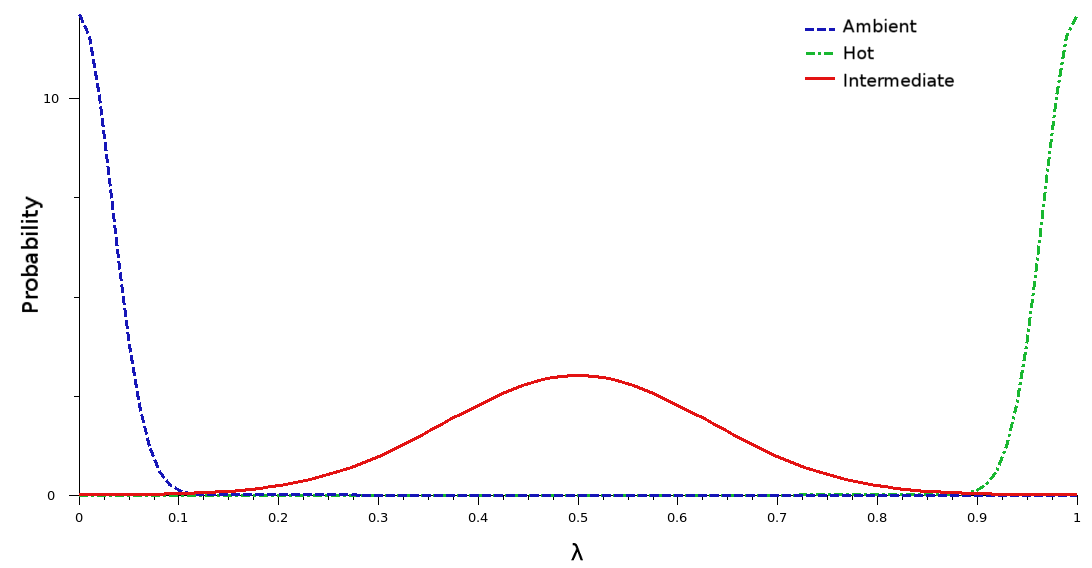}
\end{center}
\caption{Probability laws of $\lambda$ for the different operationnal mode.}
\label{fig:gauss}
\end{figure}

The final choice of $\lambda$ is done by random drawings according to gaussian laws (or half-gaussian at ends, see Figure \ref{fig:gauss}). The standard deviation of each gaussian is choosen in order to have $99\%$ of its surface inside the intervals considered above. This is equivalent to $3\sigma_{TA}=3\sigma_{TF}=0.1$ and $3\sigma_{TI}=0.4$, this give us $\sigma_{TA}=\sigma_{TF}=0.33$ and $\sigma_{TI}=0.133$ respectively. Then the different laws are given by equations (\ref{eq:pta}), (\ref{eq:ptf}) and (\ref{eq:pti}) (for all $\lambda$ taken in the previous intervals).

\begin{align}
P_{TA}(\lambda)&=\frac{1}{\sqrt{2\pi\sigma_{TA}^2}}\exp(-\frac{\lambda^2}{2\sigma_{TA}^2}) \label{eq:pta} \\
P_{TF}(\lambda)&=\frac{1}{\sqrt{2\pi\sigma_{TF}^2}}\exp(-\frac{(1-\lambda)^2}{2\sigma_{TF}^2}) \label{eq:ptf} \\
P_{TI}(\lambda)&=\frac{1}{\sqrt{2\pi\sigma_{TI}^2}}\exp(-\frac{(\lambda-0.5)^2}{2\sigma_{TI}^2}) \label{eq:pti}
\end{align}

By selecting different thermal configurations (for example a vehicle in standby where its engine and muffler are hot, its body, windows and tires at ambient temperature), we can generate the intermediate texture to plate on the 3D model.

\begin{figure}[t]
\begin{tabular}{cc}
\includegraphics[width=0.22\textwidth]{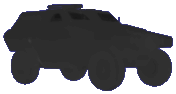} & \includegraphics[width=0.22\textwidth]{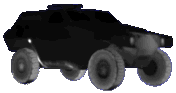} \\
\includegraphics[width=0.22\textwidth]{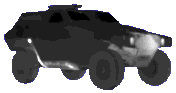} & \includegraphics[width=0.22\textwidth]{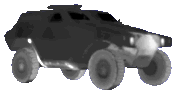}
\end{tabular}
\caption{Example of different thermal configurations generated by the proposed methode.}
\label{fig:sign}
\end{figure}

Let us present some results we get by hybrid simulation. First, Figure \ref{fig:sign} shows different thermal configurations of a same vehicle presented in the same point of view. We can see that it is possible to create realistic IR signatures which correspond to predefined operational states (vehicle completely motionless, vehicle in motion, $\ldots$). In conclusion, the method enables us the generation of all needed views of a vehicle.

\begin{figure}[!h]
\begin{center}
\includegraphics[scale=0.5]{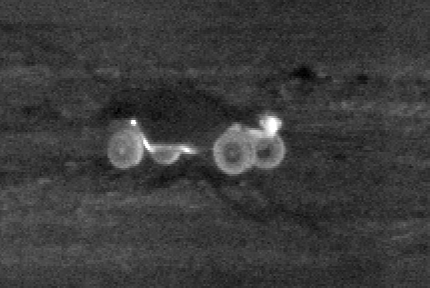} 
\includegraphics[scale=0.5]{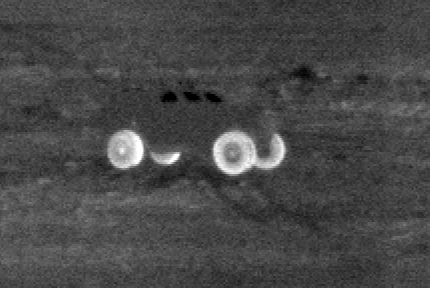} 
\includegraphics[scale=0.5]{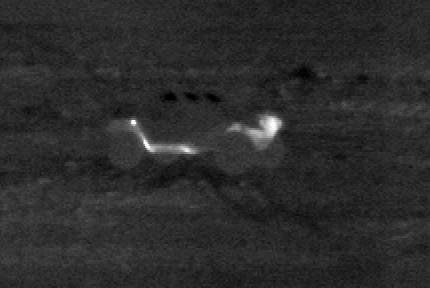} 
\caption{Example of a same scene with different thermal configurations of a vehicle.}
\label{fig:incr}
\end{center}
\end{figure}

Second, these new signatures are added to a new target database which will be used by the hybrid sce\-ne generator. This allows us to generate scenes which take into account the intrinsic thermal variability of targets by superimposing the wanted target ta\-ken in this new database. Figure \ref{fig:incr} shows an example of a same scene, generated with the same image quality constraints, containing the same vehicle with different thermal configurations.

\subsection{Sequences generation}
\begin{figure}[!h]
\begin{center}
\includegraphics[width=0.45\textwidth]{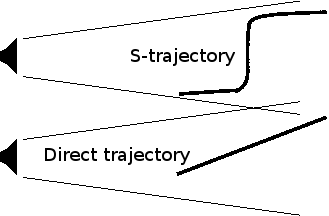}
\end{center}
\caption{The two chosen trajectories for sequence generation. Top: S trajectory, bottom: direct trajectory.}
\label{fig:traj}
\end{figure}

In order to assess the case of moving targets, we also need some sequence datasets. As in the previous described methodology, we are also interested in exhaustive sequences. Then, we want to keep the same methodology to generate the sequence data\-sets.\\
As our method doesn't use 3D terrain modelling we cannot deal with every kind of trajectories (for example, we cannot take care about ground elevation). But, as we have 3D models of targets, we can easily get the wanted direction of view of it. This will be useful because we can choose a trajectory, which is assumed to be on a plane, and from the position of the sensor we can deduce the corresponding orientation. At this time, we choose only two trajectories (see figure \ref{fig:traj}). The first one, which we called ``S-trajectory'', alterns the orientation of the target (front, next side, next front views) in order to evaluate if the orientation has an impact on the performances of the algorithm. The second trajectory is a ``Direct-trajectory'', the orientation of the target doesn't chan\-ge during the movement.\\

\begin{figure}[!h]
\begin{center}
\includegraphics[width=0.47\textwidth]{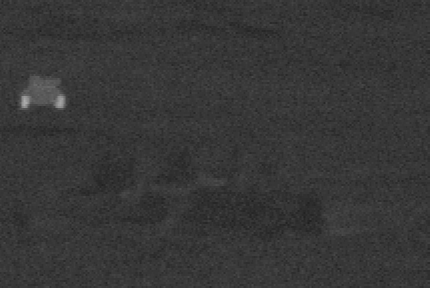} 
\includegraphics[width=0.47\textwidth]{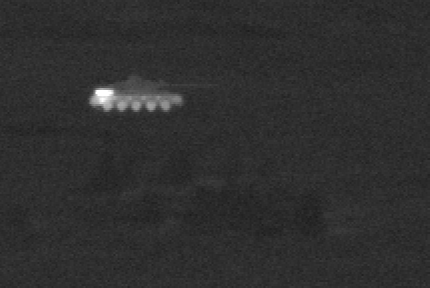} 
\includegraphics[width=0.47\textwidth]{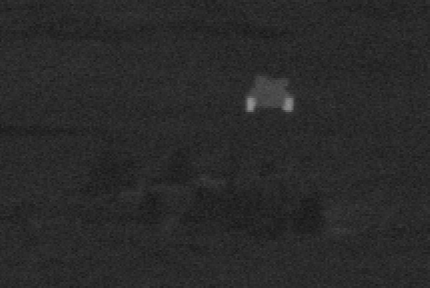} 
\caption{Example of a ``S-trajectory'' sequence for a target at 950m.}
\label{fig:seq}
\end{center}
\end{figure}

Then, for each position along the trajectory, we do the corresponding projection with the good orientation of the 3D target model. These signatures are used as inputs in the hybrid simulation. We assume that the parameters of the hybrid method are choosen for the whole sequence and does not depend on the position of the target. This permits us to generate sequences easily and create a huge dataset for ATD/R evaluation. In figure \ref{fig:seq}, we show respectively frames 400, 500 and 600 of an S-trajectory sequence of a target at 950m. We can see that we have different appearances of the target during the sequence which is important to assess the recognition behavior of algorithms.

\section{ATD/R algorithm assessments}\label{sec:test}
\begin{figure}[!h]
\begin{center}
\includegraphics[width=0.47\textwidth]{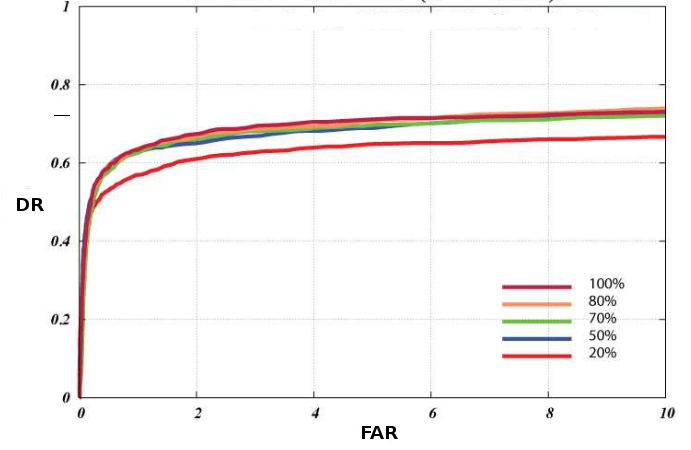} 
\includegraphics[width=0.47\textwidth]{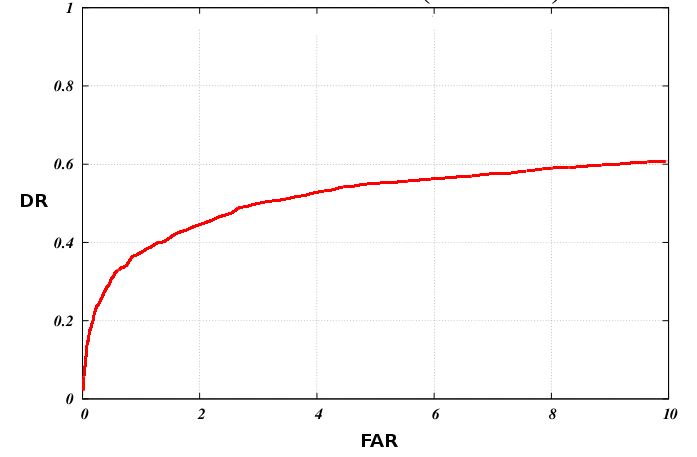} 
\includegraphics[width=0.47\textwidth]{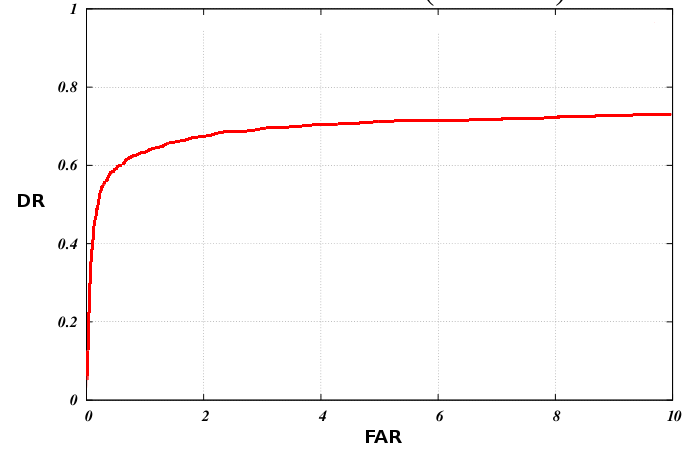} 
\caption{Example of ROC curves obtained by the detection assessment methodology.}
\label{fig:roc}
\end{center}
\end{figure}

In the program called 2ACI, different research teams work on ATD/R and Tracking algorithms. Today, some preliminary evaluations
are conducted with some algorithms. In order to have some exhaustive results, we generate a huge database which contains many scenarios corresponding to different levels of difficulties. This database was created by the method presented in section \ref{sec:data} and contains more than 37000 images. In figure \ref{fig:roc}, we present some examples of results we get by the previous described methodology for the case of detection by plotting the corresponding ROC curves. The plot on top shows the ROC curve obtained for different choice of parameters in the algorithm. This permits to see the behavior of the algorithm regarding its parameters. The next two ones show the performance of the algorithm on two different scenarios of the database. The second curve corresponds to a more complicated scenario than the curve on the bottom of the figure.

\section{Conclusion}
In this communication, we propose a complete me\-tho\-dology to assess Automatique Target Detection/ Recognition and Tracking algorithms on infrared images. We propose an original method to build huge datasets by ``hybrid'' simulation which permits a full control of image quality. This allows us to evaluate the algorithms in an exhaustive way. We also define different metrics which are representatives of the measured performances.\\

In the future, in DGA, we plan to use this metho\-dology to assess all ATD/R and Tracking algorithms. This will permit us to easily compare the algorithms proposed to DGA. \\

Finally, some current works are dedicated tp adapt the presented methodology to the visible imagery. The ATD/R performances metrics remains unchanged from the thermal case. Nevertheless, the image quality figures of merit used for the image database generation should be reviewed, because of the image formation process in the physical domain which is more complicated than in the thermal domain. Actually, the reflective part of the photonic flux coming from the target is dominant compared to the emissive part, which makes the signature appearance more variable (BRDF effects, shadow phenomenon). In addition, the visible imagers are often sensitive to colors, which increase again the diversity of the target/background combinations to consider. In spite of the added complication, the interest of this method remains valid: simplifying the apparent complexity of the image formation process in order to explore more easily and more exhaustively the combination set of the possible cases.

\bibliographystyle{plain}
\bibliography{optro2010}
\end{document}